\documentclass[12pt, preprint]{aastex}
\received{ }
\revised{ }
\accepted{ }
\cpright{AAS}{2000}
\shorttitle{HI in S0 Galaxies}
\shortauthors{Sage \& Welch}

\begin{document}

\title{The Cool ISM in Elliptical Galaxies. I. A Survey of Molecular Gas}

\author{Leslie J. Sage}
\affil{University of Maryland}
\affil{Department of Astronomy}
\affil{College Park, Maryland 20742}
\affil{USA}
\email{lsage@astro.umd.edu}
\and
\author{Gary A. Welch}
\affil{Saint Mary's University}
\affil{Department of Astronomy and Physics}
\affil{Halifax, Nova Scotia B3H 3C3}
\affil{Canada}
\email{gwelch@orion.stmarys.ca}
\and
\author{Lisa M. Young}
\affil{New Mexico Institute of Mining and Technology}
\affil{Department of Physics}
\affil{Socorro, New Mexico}
\email{lyoung@physics.nmt.edu}
    
\begin{abstract}

We present preliminary results from a survey of CO emission from members of 
a volume-limited sample of non-cluster elliptical galaxies. Our intent
is to compare the gas properties of these ellipticals to a sample of 
lenticulars selected using similar criteria. The data, 
although still sparse, suggest that the cool gas in ellipticals shows 
the same puzzling upper mass cutoff found in the lenticular galaxies.  
We find, however, significantly lower detection 
rates and possibly much lower H$_2$/HI mass ratios in the ellipticals.  
The detection rate is higher among the lower-mass galaxies, as has
been found previously.  This seems puzzling given that the
deeper potential wells of the larger galaxies ought to make gas
retention easier, but perhaps that effect is overwhelmed by feedback from the
central supermassive black hole. As we have observed $\sim 40$ percent
of our full sample, the conclusions are necessarily tentative at this
time. 

\end{abstract}

\keywords{galaxies: elliptical and lenticular, cD - galaxies:evolution -
galaxies: ISM }

\section{Introduction}

There is as yet no clear, coherent and complete explanation for the 
origins of elliptical galaxies and their gas contents.  Since the
early numerical 
simulations of Toomre \& Toomre (1972) it has been evident that mergers between
spirals can produce objects whose structure resembles that of an elliptical 
galaxy (e.g. Barnes \& Hernquist 1992; Naab \& Burkert 2003).  A
current view is that bright  
ellipticals come from early, major mergers between disks (c.f. Naab et
al. 1999; Naab \& Burkert 2003).  Minor mergers - those involving mass 
ratios of roughly 3:1 and larger - probably formed the fainter galaxies, 
those with disk components and large rotational speeds.  In all cases 
some dissipation appears to be needed (e.g. Barnes \& Hernquist 1996; 
Shioya \& Bekki 1998; Cretton et al. 2001; Robertson et al. 2006;
Naab, Jessiet \& Burkert 2006), especially for making 
fainter galaxies, but how much is not clear.  To further complicate 
the situation, it is becoming clear that mergers between gas-poor objects 
have contributed importantly to the growth of many luminous
ellipticals (van Dokkum 2005, Naab et al. 2006; Bell et al. 2006;
Boylan-Kolchin et al. 2006) -  
the so-called dry or red 
merger hypothesis.  On the other hand, the classical monolithic 
formation idea has refused to die; a modified version suggested by 
Kormendy (1989) and 
Kobayashi (2004) is still attractive for explaining radial 
gradients in absorption line strengths (Ogando et al 2005).

There is not yet even clear agreement on the observed properties of 
ellipticals, especially the properties of their interstellar media.  For example, Lees et al. (1991) and Knapp \& Rupen (1996) 
found molecular gas in about 
half of the IRAS-selected galaxies they observed, while Georgakakis et al. 
(2001) reported CO in just 25 percent (HI in almost 50 percent).  
Wiklind et al. (1995) found CO emission from 55 percent. The 
Georgakakis et al. work, however, was based on a sample of interacting 
galaxies at different stages of evolution (the stage was determined by 
the optical colours according to the prescription given in Georgakakis 
et al. 2000), while the other three studies used an IRAS criterion -- 
the 100 micron flux $S_{100}$ had to be $>1$ Jy. Wiklind et al. 
(1995) carefully assessed the morphology of each galaxy in their sample, and 
when they removed the merger candidates the detection rate (of presumably 
normal ellipticals) dropped to about 40 percent. 
									
What is clear, and has been since Faber \& Gallagher (1976), is that gas 
from evolved stars {\it must} be returned to the interstellar medium of a 
galaxy, where after a Hubble time it will (if not recycled back into stars 
or ejected in galactic winds) account for up to $\sim$10 percent of the visible
mass (i.e. Ciotti et al. 1991, Kennicutt et al. 1994, 
Brighenti \& Mathews 1997). It is also clear that new gas can 
be acquired by merging with like-sized galaxies, absorbing dwarf companions, 
or accreting pristine material from the nearby intergalactic medium.  

Simple physical arguments show that the mixing of returned gas within a 
galaxy dominated by random motions will heat that gas to temperatures of 
$10^6-10^7$ K, which explains why ellipticals with $L_{\rm B} \ge
3\times 10^{10}$ L$_{\odot}$ have hot X-ray emitting halos
(O'Sullivan, Forbes \& Ponman 2001).  Continuing energy input from aging stars (supernovae, novae and
winds from very massive stars) keeps at least a portion of that gas hot, and 
may even blow it out of the galaxy.  The details of energy transfer to the ISM,
though, are complex and the general outcome correspondingly uncertain (c.f. 
Mathews 1990).  Simulations show that significant amounts of gas can cool near 
the centers (e.g. Brighenti \& Mathews 1997; Pellegrini \& Ciotti 1998).
     
Efforts to follow the evolution of the hot ISM in elliptical galaxies have led
to the strong prediction that only massive galaxies will develop central 
reservoirs of cool gas.  It is an ongoing puzzle, therefore, that observers 
have reported higher HI and CO detection rates among less luminous galaxies 
(Lake \& Schommer 1984, Lees et al. 1991 for HI and CO, respectively).  Lees et 
al. point to a possible bias against detecting broad, faint emission lines in 
massive objects.  Another possibility is that ISM reheating by a central AGN is
more effective in massive galaxies; for example, di Matteo et al. (2005) found 
in simulations that when the mass of the central black hole exceeds 
$\sim 10^7$ M$_{\odot}$ that the outflows generated can turn off star
formation. 

A variety of mechanisms, such as ram pressure stripping, galaxy-galaxy
interactions, and perhaps thermal interactions with the hot gas or
sputtering and grain destruction, affect the ISM of galaxies moving
within a cluster; 
field galaxies should be free of such complications, and are 
therefore the obvious starting points for understanding internal processes 
governing ISM evolution.  Published studies, though, have been biased not only 
towards cluster membership but also in favor of objects likely to contain a 
detectable ISM (e.g. far-infrared detection, optical dust features).
Following 
the philosophy of our recent study of S0 galaxies (Welch \& Sage
2003), we have 
defined a volume-limited sample of non-cluster elliptical 
galaxies with the aim to determine systematically their cool gas
contents. We used integration times designed to reveal (at the
5$\sigma$ level) just one percent of the mass expected to be returned
by stellar evolution (Ciotti et al. 1991), so that even non-detections give us physically
meaningful constraints.  We also take advantage of increased bandwidth compared to earlier surveys, which facilitates detection of rotationally broadened lines.  This 
paper reports our first results: CO observations of 18, of which we have 
detected emission (in either the 2-1 or 1-0 lines) from six. 

\section{Observations and Data Reduction}

We used the Nearby Galaxies Catalog (Tully 1988) to define a volume-limited 
sample of galaxies -- with distance $d< 25$ Mpc, and declination $>-20\deg$ -- 
excluding members of the Virgo and Fornax clusters.  We added from the RC3 (de Vaucouleurs et al. 1991) any ellipticals satisfying the same distance and declination 
criteria that were not in the Nearby Galaxies Catalog, arriving at a final 
list of 46 galaxies. Some galaxies classified as elliptical (Type -5) in the Nearby Galaxies Catalog are assigned type S0 in the RC3 (Table 1). The present sample, then, is chosen with basically the 
same criteria as in our recent work on lenticulars (Welch 
\& Sage 2003; Sage \& Welch 2006) with one exception: in the S0 study we excluded cases where interaction was obvious on the Palomar Sky Survey prints and in the present case we have retained them.  The galaxies thereby retained are NGC 3226, NGC 3640, and NGC 7464; we have not yet observed these objects.  

The entire sample is listed in Table 1. All positions and systemic 
velocities are from the NASA Extragalactic Database.\footnote{The NASA/IPAC 
Extragalactic Database (NED) is operated by the Jet Propulsion Laboratory, 
California Institute of Technology, under contract with the National 
Aeronautics and Space Administration.} 

The data were obtained through pooled (service) observations (27 October 
2004 to 8 
November 2004) and during a run from 10 May 2005 to 13 May 2005 (LJS and GAW  
present), 
at the IRAM 30m telescope in Spain. 
All data reduction was done using CLASS. The observed galaxies, along with 
CO integrated line intensities, uncertainties and line windows, are listed 
in Table 2. 

The data were co-added and reduced in the normal way, with a few obviously bad 
scans omitted. The only exception was Maffei 1, which is both large and 
affected by local (Milky Way) emission. We observed 5 points in Maffei 1 -- 
the center and $12''$ away on either side of both the major and minor axes. 
Using the data from the survey of Milky Way emission (Heyer et al. 1998) we 
identified the contamination at all positions and removed it. No residual 
emission was evident in any of the scans, so we co-added all positions. There 
was still no evidence for any emission.     
  
\section{Results}

Of the galaxies observed, we detected (at the $\ge 3\sigma$ level) six in the 
1-0 line, and three in the 2-1 line. There were no 2-1 data for Maffei 1 and 
NGC 7468 because a cable was unplugged during the service observing. 

Unlike our survey of lenticulars, the detection rate is rather low even though 
integration times were designed to detect as little as 1\% of the expected gas 
mass.  Furthermore, the 
2-1 line is much more difficult to see. There were several cases in the 
lenticular study where only the 2-1 line was seen, from which we concluded 
that the gas was very centrally concentrated (Welch \& Sage 2006). For the 
elliptical galaxies the reverse seems to be true -- the gas does  not 
seem to be particularly concentrated, based upon the limited line
information, or it is much colder than in lenticulars. The reasoning
goes as follows -- if the gas is spread over a region the size of the
1-0 beam (or larger), and not unusually warm, then the 1-0 line will appear 
stronger than the
2-1, simply because more molecules are contributing to the
emission. The lack of concentration is evident in interferometric maps
(Young 2002, 2005), where of the seven galaxies studied the CO
only one had a CO-emitting region smaller than the 1-0 beam at the 30m
telescope. 

Elliptical galaxies are known to be generally more difficult to detect in
CO emission than S0s (e.g. Lees et al. 1991).  We find CO in 78 percent of
our volume-limited S0 sample but in only 33 percent of the present,
preliminary elliptical sample.  The two samples span similar ranges of 
luminosity and local galaxy density. On the other hand, our observations so 
far have been weighted towards the more massive galaxies (we have observed 58 percent of targets brighter than M(B) = -20 but only 32 percent of the fainter ones), while it is known 
that less massive ellipticals are more likely to contain gas (i.e. Lake \& 
Schommer 1984, Lees et al. 1991)

Table 3 contains the 2-1/1-0 line ratios for the three galaxies for which we 
have $>3\sigma$ detections in both lines. The number is too small, and the 
scatter too large, for us to 
be able to conclude anything from those ratios. 

We have found published CO observations for 6 galaxies listed in Table 2: 
NGC 720 and NGC 4636 (Braine et al. 1997), NGC 2768 and NGC 7468 
(Wiklind et al. 1995), 
NGC 4697 (Knapp \& Rupen 1996), and NGC 4742 
(Lees et al. 1991).  Our results are in reasonably good agreement,
although we 
typically report somewhat lower integrated intensities or more sensitive 
upper limits.

\section{CO Detection Rates in E and SO galaxies}

The striking difference in CO detection rates between ellipticals
and lenticulars may offer important guidance for future investigations of
ISM evolution in early type galaxies.  We have speculated (Sage \& Welch 2006) that the CO emission from S0
galaxies comes mainly from gas which has cooled out of the hot, X-ray
phase - i.e.  gas returned by the stars.  It is not yet evident that the same is true in
elliptical galaxies (one of our goals is to address this issue with more
data).  Monolithic models of ISM evolution in early type galaxies incorporate a variety of factors which influence whether, and how much, gas cools from the hot 
phase (e.g. Mathews 1990, Ciotti et al. 1991, Brighenti and Mathews 1997, 
Pellegrini \& Ciotti 1998); those include the mass of the dark halo, the 
effectiveness of (type Ia) supernova heating, and AGN feedback.

More massive dark matter halos promote the cooling of gas near the center by
inhibiting the development of a galactic wind; since more hot gas is retained 
globally though, the X-ray luminosities are also higher.  
Models with high L(X) might therefore also develop central clouds of 
atomic and/or molecular gas (but those models - e.g. Ciotti et al. 1991 - do not follow the
detailed evolution of the cold ISM). In general, ellipticals are found to have 
higher X-ray luminosities and larger L(X)/L(B) 
than lenticulars (Eskridge et al. 1995a,b), consistent with them being 
more dark matter dominated.  If the cold gas in both galaxy types primarily originates in cooling from a hot phase, one might therefore expect higher CO detection 
rates in our elliptical sample, which is contrary to what we find.  We
caution, 
though, that our samples of E and S0 galaxies include relatively fewer high 
luminosity objects than the Einstein sample (with correspondingly low X-ray 
detection rates), and may therefore not reflect the trends found by 
Eskridge et al.   On the other hand, Braine et al. (1997) found no CO 
emission from a sample of six luminous ellipticals. 

AGN feedback is currently thought theoretically to be important in 
suppressing cooling
flows in both clusters and individual elliptical galaxies (di Mateo et
al. 2005; Sazonov et al. 2005), and observational evidence
supporting that view has recently been reported (Schawinski et
al. 2006).  Although the
process is complex and only partly understood (cf Brighenti \& Mathews 2002, 
2003), the current picture of AGN  
energy production (Binney \& Tabor 1995; Ciotti \& Ostriker 1997, and
more recently Brighenti \& Mathews 2006) suggests that feedback is 
likely to occur, at least 
initially, along the spin axis of the central supermassive black
hole. 
Since that axis would not often point towards the disk of a lenticular 
host galaxy, cooling gas from its disk stars might often escape being
reheated, which might lead to the observed higher detection frequency in 
S0s.  In other words, AGN reheating might be more effective in a pure 
spheroid simply because gas return is more isotropic than in a disk galaxy.    
Published simulations, however, have not yet featured directed AGN feedback 
within disks, so that explanation remains speculative. Sazonov et
al. (2005) did show that when the mass of gas drops below one percent
of the mass of stars in the central region, radiative heating by the
AGN overwhelms cooling, and much of the remaining gas will be
expelled. 

The flattened shapes of lenticulars are expected to make it easier for 
supernovae to heat and eject returned gas (c.f. D'Ercole \& Ciotti
1998).  While 
that might help explain the generally lower X-ray luminosities among S0 
galaxies, it would also seem to predict lower, not higher, CO detection 
rates among S0s, contrary to the observations.

In the hierarchical formation picture lenticular and elliptical galaxies 
are products of different merging histories.  We have already commented 
on our current, blurry picture of elliptical galaxy formation.  Even less 
obvious is the way to make lenticular galaxies in low-density environments 
(Stocke et al. 2004), although it seems that most S0s cannot simply be 
former spirals (Burstein et al. 2005).  Furthermore, it remains to
work out how differing assembly sequences for Es and S0s can account for 
differences in cold gas detection rates, in the distribution of the molecular and 
atomic phases, and possibly also in the global values of M(H$_2$)/M(HI)
(see below); presently there is much room for speculation.  For example, 
perhaps ellipticals typically form in somewhat dryer mergers than
lenticulars, as larger gas contents do seem to produce diskier
remnants. Barnes (2002) showed that, depending on merger geometry,
some tens of percent of the gas carried into a major merger could end
up in a large-scale gas disk that could later form a stellar
disk. Naab, Jesseit \& Burkert (2006) also showed that the presence of
even small amounts of gas in a merger tends to make the remnant more 
axisymmetric, more dominated by minor axis tube orbits, and less boxy 
(and so more like a lenticular galaxy). On the other hand, perhaps timing is 
important, in that ellipticals usually acquired their gas earlier 
than S0s, and have had more time to dispose of it by feeding an AGN 
and/or making stars.

Previous surveys have reported higher cool gas detection rates among 
low luminosity E and S0 galaxies (Lake \& Schommer 1984, and Lees et 
al. 1991 for HI and CO, respectively); our results show a similar 
trend, albeit from small numbers.  We find CO in only 1 of 7 galaxies 
(or 14 percent) brighter than M(B) = -20, but in 5 of 11 (45 percent) 
of fainter galaxies.  For a cut at M(B) = -19 the results are 20 
percent and 50 percent, respectively. The lower CO content in the more
massive galaxies may arise because the high-luminosity galaxies are
more likely formed in dry mergers (Bell et al. 2006; Boylan-Kolchin et
al. 2006), or perhaps the evolution of the gas is dominated by AGN
feedback (Sazonov et al. 2005). In contrast, other published 
simulations of ISM evolution inside isolated galaxies (e.g. 
Ciotti et al. 1991; Pellegrini \& Ciotti 1998) suggest that cool 
gas more readily accumulates in brighter, not fainter, ellipticals.  

In summary, the detection rates of molecular gas among early type 
galaxies are not easy to understand from their other properties, 
and useful predictions are not available from either of the two 
competing paradigms of galaxy formation.  Much more work is needed 
to bring a consensus into view.

\section{Cool Gas in the Volume Limited Sample - a First Look}

One of the most puzzling and, we believe, potentially most significant 
discoveries from our survey of S0 galaxies has been the existence of a cutoff
in the amount of cool gas present (Sage \& Welch 2006).  Regardless of 
luminosity, the most gas-rich lenticulars have $\sim$10 percent of the amount 
returned by their stars after the first 0.5 Gyr.  Table 4 presents the H$_2$ 
and HI masses for ellipticals from the present work and the literature, respectively.  Although 
we have both CO and HI data for less than half of the elliptical sample (lack 
of CO observations is the main deficiency) there is already an
indication that 
the same cutoff may be present (Figure 2);   The outstanding exception is 
NGC 7468 (Markarian 314), which has almost as much gas as predicted by Ciotti et al. (1991) - most of  
this is HI.  Although isolated, its optical structure is clearly peculiar.  
Perhaps NGC 7468 has recently acquired additional atomic gas from nearby.

The cool gas in ellipticals and S0s may have another common attribute: 
galaxies with the most gas also have relatively more in the atomic phase.  
The correlation, shown in Figure 3, is merely suggestive due to the paucity of 
observations; the reader should compare with the analogous plot for S0 
galaxies in Figure 13 in Sage \& Welch (2006).  An interesting difference 
between the two plots is their relative offset in M(H$_2$)/M(HI): Elliptical galaxies seem more deficient in molecular gas by an order of 
magnitude than S0s, when compared at the same value of M(ISM)/M(PRE).  
If cooling flows generate most of the observed molecular gas in early-type galaxies, that result would suggest that some mechanism, perhaps AGN reheating, moderates them more effectively in ellipticals 

In summary, we have begun to assess the amount of cool gas 
within the members of a volume limited sample of elliptical galaxies.  CO is 
much more frequently detected in S0s, which we speculate might be because 
AGN feedback is less effective at reheating the gas returned within disks.   
Presently available data indicate that whatever mechanisms operate to impose 
a cutoff on the cool gas mass in S0 galaxies also operate in ellipticals.  A 
clear but still poorly established trend that more gas rich ellipticals have 
relatively more atomic gas mimics the more robust trend shown by S0s.   We 
emphasize that additional CO observations (and in some cases HI
observations) are needed to improve the
statistics, because of our small sample size. Differences in the gas
properties of ellipticals and lenticulars may reveal robust new
indications of whether their formation mechanisms differed and how
their evolution proceeded. We are not yet able to identify the physical causes of the trends we report.  More realistic simulations of ISM
evolution, based on each of the competing paradigms of galaxy evolution - monolithic and hierarchical - will be needed to accomplish that. 

This work has been supported by a Discovery Grant to GAW from the Natural 
Sciences and Engineering Council of Canada.


\pagebreak

\pagebreak
\begin{deluxetable}{lcccccc}
\tablecolumns{7}
\tablewidth{0pc}
\tabletypesize{\footnotesize}
\tablecaption{Properties of Galaxies in the Volume-limited Sample}
\tablehead{
\colhead{Name} & \colhead{R.A.} & \colhead{Dec.} & \colhead{$v_{\rm hel}$} &
\colhead{Type} & \colhead{D} & \colhead{$B_{\rm T}^{\rm o}$} \\
\colhead{} & \colhead{J2000} & \colhead{J2000} & \colhead{km s$^{-1}$} &
\colhead{} & \colhead{Mpc} & \colhead{}  
}
\startdata

NGC 584  & 01:31:20.7  & -06:52:05 & 1802   & E4 & 23.4 & -20.68\\
NGC 596 & 01:31:36.8 &  -06:53:37 & 1876 & E+pec(unc)   &  23.8 &  -20.22\\
NGC 636 & 01:39:06.5 &  -07:30:45 & 1860 & E3   &24.2 & -19.71\\
NGC 720 & 01:53:00.5 &  -13:44:19 & 1745 & E5   & 20.3 & -20.38\\
NGC 821 & 02:08:21.1 &  +10:59:42 & 1735 & E6(doubtful) & 23.2  & -20.21\\\\
                                                
NGC 855 & 02:14:03.6 &  +27:52:38 & 595 & E &   4.3 &   -15.28\\
IC 225  & 02:23:53.8 &  +01:09:38 & 1535 & E &  17.0 &  -16.72\\
Maffei 1 & 02:36:35.4 & +59:39:19 & 13  & gE &  3.6  &  -20.50\\
NGC 1052 & 02:41:04.8 & -08:15:21 & 1510 & E4 & 17.8    & -19.81\\
NGC 1172 & 03:01:36.0 & -14:50:12 & 1669 & E+(unc) & 18.3 & -18.50\\\\
                                                        
NGC 1297 & 03:19:14.2 & -19:06:00 & 1578 & SAB0(pec)  & 18.3  & -18.71\\
Haro 20/UGCA 073 & 03:28:14.5  & -17:25:10. & 1866   & E+(doubtful) &   21.9  &
-16.94\\
NGC 1407   & 03:40:11.9 & -18:34:49 & 1779  &  E0  & 21.6 & -21.02\\
NGC 2768 & 09:11:37.5 & +60:02:14 & 1373  & E6(unc) &   23.7  & -21.13\\
NGC 3073 & 10:00:52.1 & +55:37:08 & 1155 & SAB0- & 19.3 & -18.03\\\\
                                                
N3115 DW1  & 10:05:41.6 & -07:58:53.4 & 698  &  "dE1,N" & 13.4  & -17.07\\
NGC 3156  & 10:12:41.2 & +03:07:46 & 1318  &    S0(unc) & 18.6  & -18.54\\
NGC 3193 & 10:18:24.9 & +21:53:55 & 1399 & E2 & 23.2 &  -20.05\\
NGC 3226 & 10:23:27.0 & +19:53:55 & 1151 & E3 pec(unc) & 23.4 & -19.57\\
NGC 3377 & 10:47:49.6 & +13:59:08 & 665 & E5+ & 8.1  &  -18.55\\\\
                                                
NGC 3379 & 10:47:49.6 & +12:34:54 &  911  & E1 & 8.1  & -19.39\\
UGC 5955 & 10:52:04.2 & +71:46:23 & 1249 & E &  16.8  & -17.12\\
NGC 3522 & 11:06:40.4 & +20:05:08  & 1221 & E &         20.6  & -17.46\\
IC 678  & 11:14:06.4  & +06:34:37 & 968  & E &  17.6 &  -16.48\\
NGC 3605 & 11:16:46.6 & +18:01:02 & 668  & E4   &  16.8 & -18.03\\\\
                                                
NGC 3608 & 11:16:58.9 & +18:08:55 & 1253 & E2 & 23.4  & -19.94\\
NGC 3640 & 11:21:06.8 & +03:14:05 & 1251 & E3 & 24.2 &  -20.78\\
NGC 3818 & 11:41:57.3 & -06:09:20 & 1701 & E5 & 24.7 &  -19.52\\
NGC 4033 & 12:00:34.7 & -17:50:33 & 1617 & E6 & 23.9 &  -19.49\\
NGC 4125 & 12:08:06.0 & +65:10:27 & 1356 & E6 pec & 24.2 & -21.35\\\\

NGC 4239 & 12:17:14.9 & +16:31:53 & 940 & E & 16.6  &   -18.09\\
UGC 7354 & 12:19:09.9 & +03:51:21 & 1526 & E pec(unc) & 14.6 &  -16.52\\
NGC 4278 & 12:20:06.8 & +29:16:51 & 649 & E1+ & 9.7  &  -18.82\\
NGC 4283 &  12:20:20.8 & +29:18:39 & 984 & E0 & 9.7  &  -17.02\\
NGC 4308 &  12:21:56.9 & +30:04:27 & 589 & E(unc) & 9.7 & -15.83\\\\

NGC 4494 &  12:31:24.0  & +25:46:30 & 1344 & E1+ & 9.7  & -19.23\\
UGC 7767 & 12:35:32.4 & +73:40:29 & 1282 & E &  17.9  & -17.73\\
NGC 4648 & 12:41:44.4 & +74:25:15  & 1414 & E3  & 20.4  & -18.84\\
NGC 4627 & 12:41:59.7 & +32:34:25 & 542 & E4 pec & 13.7 & -17.71\\
NGC 4636 & 12:42:49.9 & +02:41:16 & 938 & E0+   & 17.0  & -20.68\\\\

UGCA 298 & 12:46:55.4 & +26:33:51 & 801 &  E+(unc) & 8.9 & -15.14\\
NGC 4697 & 12:48:35.9 & -05:48:03 & 1241 & E6 & 23.3 &  -21.67\\
NGC 4742 & 12:51:48.0 & -10:27:17 & 1270 & E4(unc) & 23.2 & -19.81\\
NGC 5845 & 15:06:00.8 & +01:38:02 & 1450 & E (unc) & 21.9 & -18.16\\
NGC 7464 & 23:01:53.7 & +15:58:26 & 1875 & E1 pec(unc)  & 19.2  & -17.39\\\\

NGC 7468 & 23:02:59.2 & +16:36:19 & 2081 & E3 pec(unc)  & 23.2  & -18.03\\

\enddata

\tablecomments{Columns contain galaxy name, coordinates at epoch 2000,
heliocentric radial velocity, morphological type from the Third Reference Catalog of Bright Galaxies (RC3), distance in Mpc, either from the Tully Catalog or, when unavailable, from V$_{3K}$ in the RC3 with H$_0$=75 km sec$^{-1}$ Mpc$^{-1}$, total corrected blue apparent magnitude from the RC3.}
\end{deluxetable}

\begin{deluxetable}{lccccc}
\tablewidth{0pc}
\tabletypesize{\footnotesize}
\tablecolumns{6}
\tablecaption{Integrated Intensities}
\tablehead{
\colhead{Name} & \colhead{Window} & \colhead{I$_{CO}$(1-0)} & 
\colhead{rms} &  \colhead{I$_{CO}$(2-1)} & \colhead{rms} \\
\colhead{} & \colhead{km/s} &
\colhead{K km s$^{-1}$} & \colhead{K}  & \colhead{K km s$^{-1}$} 
& \colhead{K} 
}
\startdata

NGC 636 & 1773-1931 & $0.34\pm 0.23$ & 0.0051 & $<0.27$ & 0.0060 \\
NGC 720 & 1543-1972 & $0.63\pm 0.47$ & 0.0054 & $<0.46$ & 0.0054 \\
IC 225 & 1523-1594 & $0.21\pm 0.046$  & 0.0016 & $0.16\pm 0.033$ & 0.0012 \\
Maffei 1 & -99-166 & $<0.13$  & 0.0022  &   &  \\
NGC 1407 & 1478-2058 & $<0.58$ & 0.0051 &  $<0.62$ & 0.0055 \\\\

NGC 2768 & 1168-1571 & $0.67\pm 0.28$ & 0.0035 & $0.92\pm 0.30$ & 0.0038  \\
NGC 3073 & 1093-1205 & $0.50\pm 0.083$ & 0.0023 & $0.45\pm 0.083$ & 0.0023 \\
NGC 3115 DW1 &  517-901 & $<0.26$ & 0.0034 & $<0.19$ & 0.0025 \\
NGC 3193 & 1203-1602 & $<0.67$ & 0.0085 & $<0.87$ & 0.011  \\
NGC 3605 &  425-907  & $<0.41$ & 0.0046 & $<0.38$ & 0.0042  \\\\

NGC 4239 & 768-1091  & $0.46\pm 0.29$ & 0.0043 & $<0.34$ & 0.0050  \\
NGC 4283 & 871-1110  & $0.88\pm 0.21$ & 0.0038 & $<0.26$ & 0.0047 \\
NGC 4494 & 1061-1541 & $1.44\pm 0.39$ & 0.0044 & $0.87\pm 0.53$ & 0.0060  \\
NGC 4648 & 1251-1672 & $<0.31$ & 0.0038 & $0.58\pm 0.30$ & 0.0037  \\\\
NGC 4636 & 794-1114 & $0.24\pm 0.20$ & 0.0029  & $<0.13$ & 0.0020  \\

NGC 4697 & 1057-1482  & $<0.44$ & 0.0052 & $0.75\pm 0.51$ & 0.0060 \\
NGC 4742 & 1057-1482 & $0.32\pm 0.21$ & 0.0027 & $<0.23$ & 0.0029  \\
NGC 7468 &  1966-2319 & $1.06\pm 0.23$ & 0.0032 & &  \\

\enddata

\tablecomments{Columns contain the galaxy name, location of line window, 
and for both CO lines the  integrated line intensity in
the line window and its formal standard deviation along with rms channel 
noise in the smoothed spectrum. All upper limits are $1\sigma$ and they are 
used whenever the formal line intensity is less than $1\sigma$. For
two 
galaxies -- Maffei 1 and NGC 7468 -- no 2-1 data were obtained because a 
cable was unplugged during the pooled observations. For Maffei 1, we observed 
5 positions -- center, and $12''$ away along both the major an minor axes. 
No emission was evident at any position. The result above contains data from 
all positions co-added, with the local Milky Way emission removed.}





\end{deluxetable}


\begin{deluxetable}{lc}
\tablecolumns{2}
\tablewidth{0pc}
\tabletypesize{\footnotesize}
\tablecaption{2-1/1-0 line ratios}
\tablehead{
\colhead{Name} &  \colhead{I$_{CO}(2-1)/{\rm I}_{CO}(1-0)$)} \\
}
\startdata

IC 225 & $0.76\pm 0.06$ \\
NGC 2768 & $1.4\pm 0.4$ \\
NGC 3073 & $0.9\pm 0.1$ \\

\enddata

\tablecomments{For the three galaxies with $>3\sigma$ detections in both 
lines we have calculated the line ratios (with all the usual caveats about 
differing beam sizes).}
\end{deluxetable}
\pagebreak
\begin{deluxetable}{lrrc}
\tablecolumns{4}
\tablewidth{0pc}
\tablecaption{Total Cool Gas Masses}
\tablehead{
\colhead{Name} & \colhead{M(H$_2$)} & \colhead{M(HI)} 
& \colhead{HI reference}  \\
\colhead{} & \colhead{($M_{\sun}$)} & \colhead{($M_{\sun}$)} & 
\colhead{}  \\  
}
\startdata
NGC 636  & $<1.67\times 10^7$ & $1.66\times 10^8$  & 1 \\
NGC 720  & $<2.40\times 10^7$ & $<2.58\times 10^8$ & 1 \\
IC 225   & $2.50\times 10^6$  &                    &  \\
Maffei 1 & $<2.08\times 10^5$ &                    &  \\
NGC 1407 & $<3.35\times 10^7$ & $<9.79\times 10^8$ & 1 \\\\

NGC 2768 &  $1.55\times 10^7$ & $1.98\times 10^8$  & 2 \\
NGC 3073 &  $7.68\times 10^6$ & $1.66\times 10^8$  & 3 \\
NGC 3115 DW 1 & $<5.78\times 10^6$ &               &   \\
NGC 3193 & $<4.46\times 10^7$ & $<7.98\times 10^7$ & 4 \\
NGC 3605 & $<1.43\times 10^7$ & $<2.39\times 10^7$ & 5 \\\\

NGC 4239 & $<9.89\times 10^6$ & $<9.12\times 10^6$ & 6 \\
NGC 4283 & $3.41\times 10^6$  & $4.15\times 10^7$  & 1 \\
NGC 4494 & $5.59\times 10^6$  & $<9.96\times 10^6$ & 7 \\
NGC 4648 & $<1.60\times 10^7$ & $<1.76\times 10^7$ & 8 \\
NGC 4636 & $<7.15\times 10^6$ & $<6.12\times 10^7$ & 9 \\\\

NGC 4697 & $<2.96\times 10^7$ & $<1.84\times 10^9$ & 10 \\
NGC 4742 & $<1.40\times 10^7$ & $<4.22\times 10^8$ & 1 \\
NGC 7468 & $2.35\times 10^7$  & $1.59\times 10^9$  & 2 \\\\

\enddata
\pagebreak
\tablecomments{Columns contain galaxy name, H$_2$ mass or upper limit from this
work, HI mass or upper limit, source of HI estimate.  A CO-to-H$_2$ conversion 
factor of $2.3\times 10^{20}$ mol. cm$^{-2}$ (K kms)$^{-1}$ has been used.  
We cannot find HI observations for 3 galaxies.  Upper limits are $3\sigma$.}




\tablerefs{         (1) \citep{huc94};  (2) \citet{huc95};
(3) \citet{irw87};  (4) \citet{wil91};  (5) \citet{kna79};  
(6) \citet{ls84};
   (7) \citet{brg92};  (8) \citet{ric87};  
(9) \citet{kt83};   (10) \citet{gal75}.  }
\end{deluxetable}
%
%
%
\begin{figure}
\figurenum{1}
\includegraphics[scale=0.6]{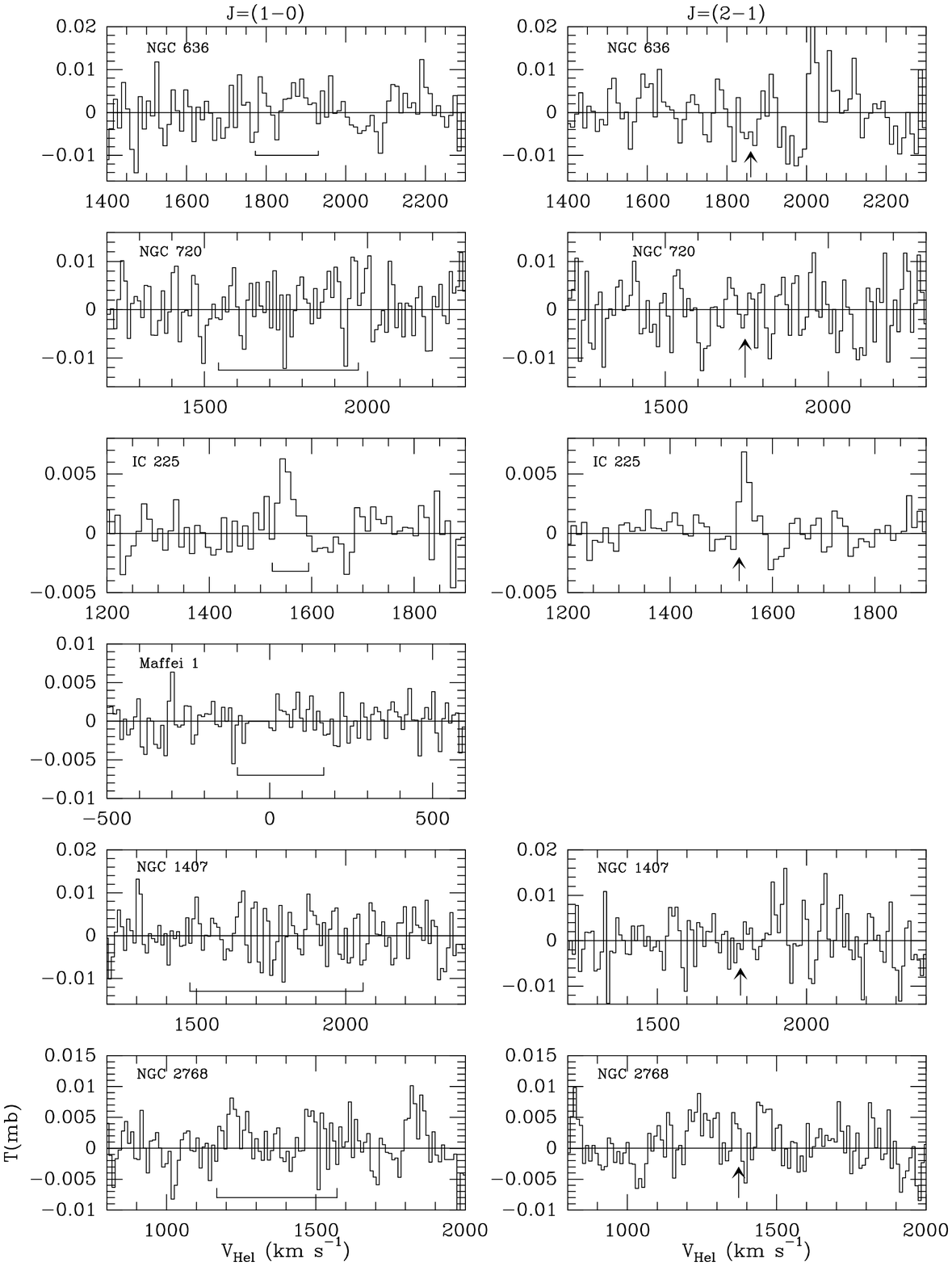}
\caption{CO spectra from the IRAM 30m telescope. The arrow in the
  J=2-1 column indicates 
the optical systemic velocity from NED. The solid line in the J=1-0
column shows the 
velocity range over 
which the integrated line intensity was calculated. }
\end{figure}

\begin{figure}
\figurenum{1}
\plotone{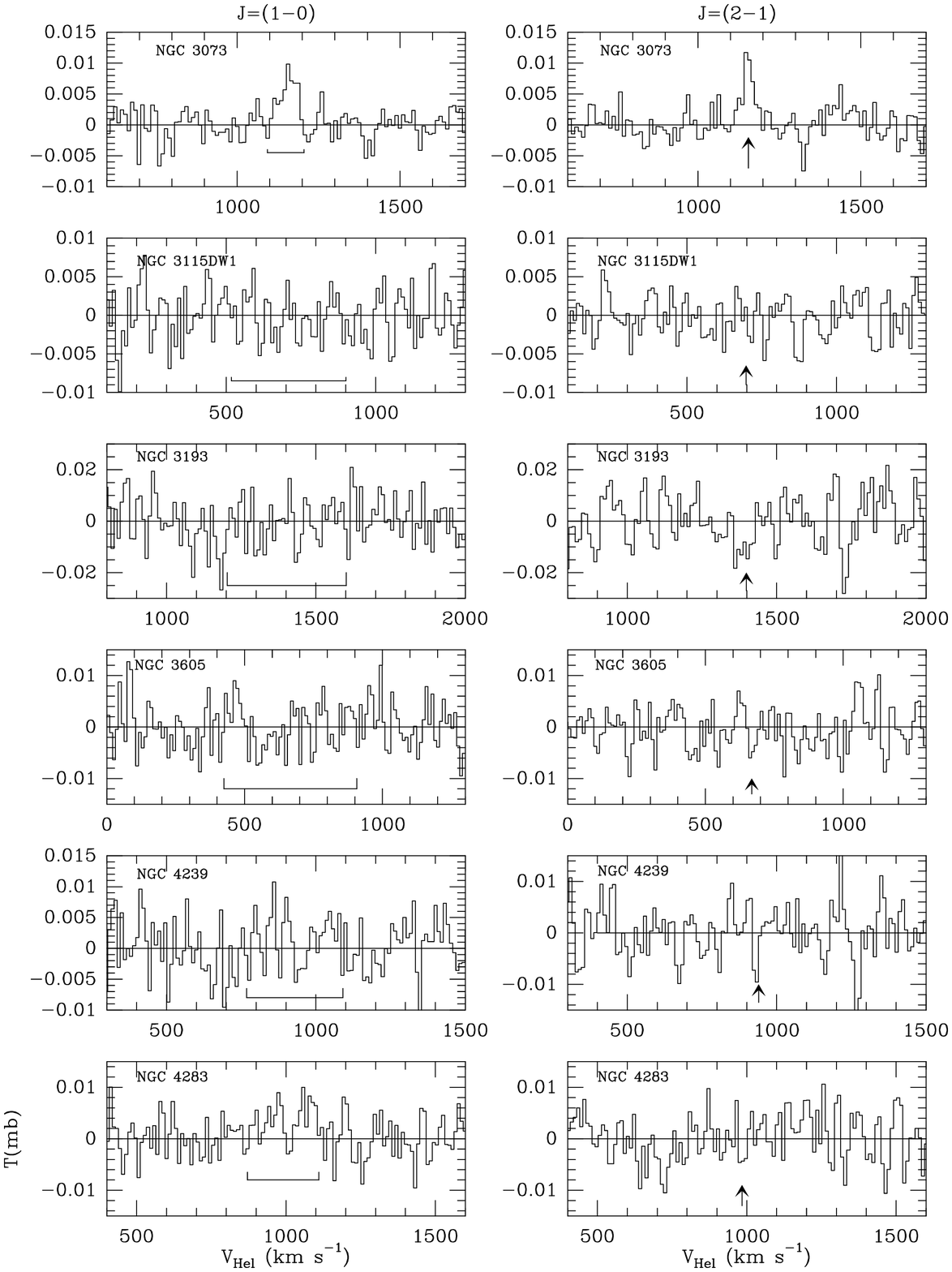}
\end{figure}

\begin{figure}
\figurenum{1}
\plotone{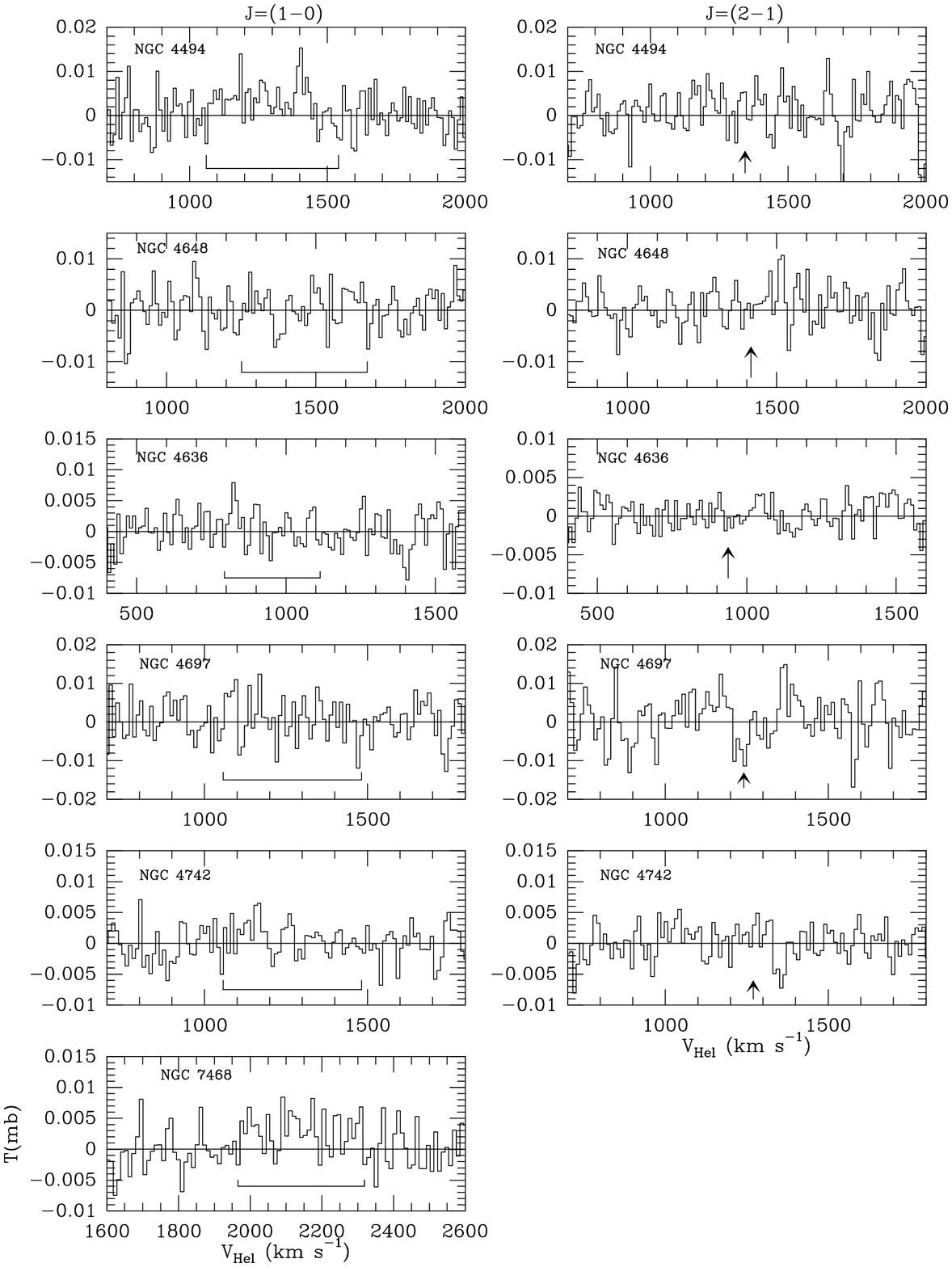}
\end{figure}

%
%
\begin{figure}
\figurenum{2}
\includegraphics[scale=0.5, angle=-90]{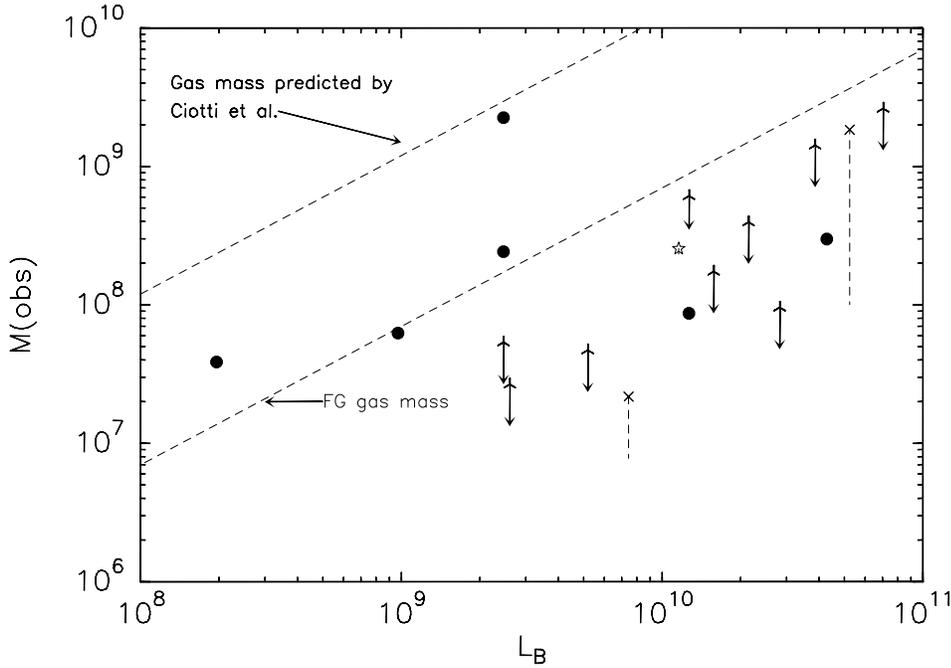}
\caption{The observed gas mass as a function of blue luminosity, compared to the mass predicted to be returned by evolving stars (lines).  Filled
circles are derived from sums of measured HI and H$_2$
masses. Measurements of M(H$_2$)  added to $3\sigma$ upper limits on
M(HI) are shown as "x", with dashed lines 
extending down to the value of M(H$_2$).  The star indicates an analogous 
treatment of a measurement of M(HI) and a $3\sigma$ limit on M(H$_2$).  
Three-armed crosses with downward arrows mark the sums of two $3\sigma$ mass 
limits.  All values are scaled by a factor of 1.4 to account for Helium. }
\end{figure}

\begin{figure}
\figurenum{3}
\includegraphics[scale=0.5, angle=-90]{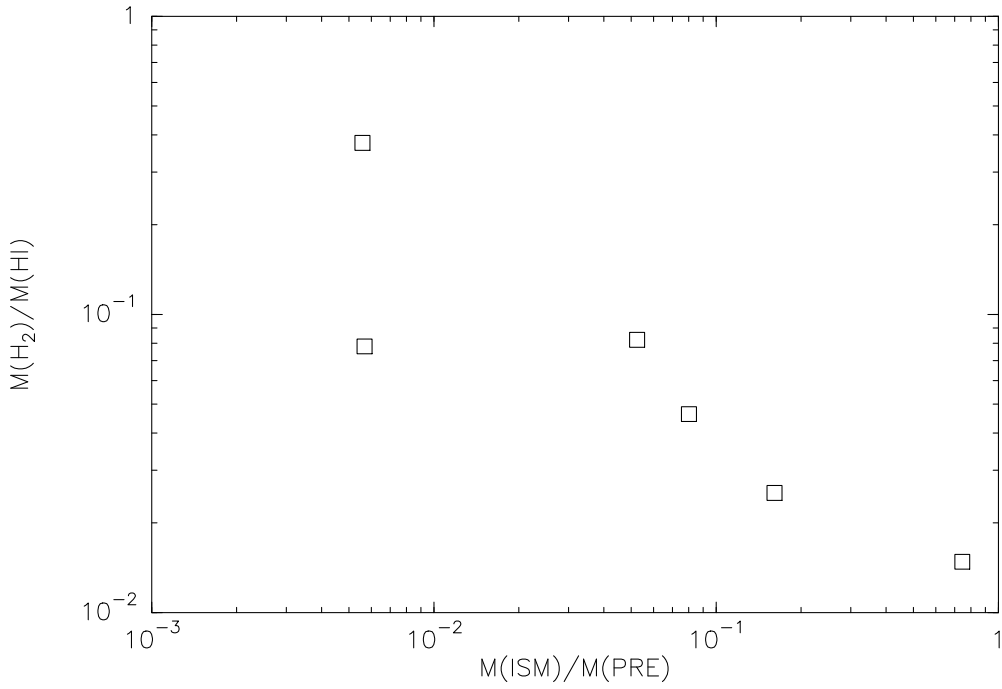}
\caption{The ratio of molecular to atomic gas mass as a function of
  the mass fraction, i.e. the ratio of the total observed mass of cool gas
  to the mass predicted by the Ciotti et al. analytical approximation - 
  compare to Figure 13 in Sage \& Welch 2006 for S0 galaxies.} 
\end{figure}

\end{document}